\documentclass{ieeeaccess}
\usepackage{cite}
\usepackage{amsmath,amssymb,amsfonts}
\usepackage{algorithmic}
\usepackage{graphicx}
\usepackage{textcomp}

\def\BibTeX{{\rm B\kern-.05em{\sc i\kern-.025em b}\kern-.08em
   T\kern-.1667em\lower.7ex\hbox{E}\kern-.125emX}}

\begin{document}
\history{Date of publication xxxx 00, 0000, date of current version xxxx 00, 0000.}
\doi{10.1109/ACCESS.2023.0322000}

\title{Automatic modulation classification for MIMO system based on the mutual information feature extraction}
\author{\uppercase{N. Ussipov}\authorrefmark{1}, 
\uppercase {S. Akhtanov} \authorrefmark{1},
\uppercase{Z. Zhanabaev}\authorrefmark{1},
\uppercase {D. Turlykozhayeva} \authorrefmark{1},
\uppercase {B. Karibayev} \authorrefmark{2},
\uppercase {T. Namazbayev} \authorrefmark{1}, 
\uppercase {D. Almen} \authorrefmark{1}, 
\uppercase {A. Akhmetali} \authorrefmark{1},
\uppercase {X. Tang} \authorrefmark{3}, (Member, IEEE) }

\address[1]{Faculty of Physics and Technology, Al-Farabi Kazakh National University, 71 al-Farabi Ave., Almaty, Kazakhstan}
\address[2]{Department of Telecommunication Engineering, Almaty University of Power Engineering and Telecommunications named after Gumarbek Daukeyev, Baytursynuli 126/1, Almaty, Kazakhstan}
\address[3]{School of Electronics and Information, Northwestern Polytechinical University, 127 West Youyi Road, Xi'an, China}

\markboth
{Author \headeretal: Preparation of Papers for IEEE TRANSACTIONS and JOURNALS}
{Author \headeretal: Preparation of Papers for IEEE TRANSACTIONS and JOURNALS}

\corresp{Corresponding author: D. Turlykozhayeva (e-mail: dana.turlykozhayeva@kaznu.edu.kz)}

\begin{abstract}
Automatic Modulation Classification (AMC) is an essential technology that is widely applied into various communications scenarios. In recent years, many Machine Learning and Deep-Learning methods have been introduced into AMC, and a lot of them apply different approaches to eliminate interference in complex Multiple-Input and Multiple-Output (MIMO) signals and improve classification performance. However, in practical communication systems, the perfect elimination of MIMO signal interference is impossible, and therefore classification performance suffers. In this paper, we propose a new AMC algorithm for MIMO system based on mutual information (MI) features extraction, which does not require a large amount of training data and the elimination of MIMO signal interference. In this approach, features based on mutual information are extracted using In-Phase and Quadrature (IQ) constellation diagrams of MIMO signals, which have not been explored previously. Our method can be effective since mutual information considers the interdependencies among variables and measures how much information about one variable reduces uncertainty about another, providing a valuable perspective for extracting higher-level and interesting features from the data. The effectiveness of our method is evaluated on several model and real-world datasets, and its applicability is proven.
\end{abstract}


\titlepgskip=-21pt

\maketitle

\section{Introduction}
\label{sec:introduction}
\PARstart{A}{utomatic} Modulation Classification (AMC) is an important technology that provides modulation information of the incoming radio signals and plays a significant role in practical civilian and military applications such as cognitive radio, interference recognition, spectrum management, data encryption, etc. AMC has gained considerable research interest in recent years since it aims to automatically identify the modulation types of wireless communications signals without prior information \cite{Meng1, Wang2, Zhou3, Dileep4, Lee5}. Among the various modulation types, quadrature amplitude modulation (QAM) and phase-shift keying (PSK) find widespread application in communication systems. Additionally, when compared to other modulation types, classifying higher-order QAM poses a greater challenge \cite{Zhu6}. Hence, this study specifically concentrates on high order 16QAM and 64QAM modulations.

Many modulation classification algorithms have been proposed for both single-input and single-output (SISO) systems and MIMO systems and these methods are usually classified into two major groups: likelihood-based (LB) and feature-based (FB). LB methods are based on the likelihood function and are considered optimal but generally complex to implement \cite{Xu7, Ozdemir8}. FB methods depend on feature extraction and classifier design and compared to LB methods are more applicable in practical implementations and more popular in AMC due to the ease of implementation \cite{Majhi9, Al-Nuaimi10}. Various types of features, such as instantaneous time-domain features, transform-domain features, and statistical features have been investigated and used in AMC algorithms \cite{Hazza11, Hong12}. After extracting the features, the classifier is used to determine the class of modulated signals. Until now, various types of classifiers have been used in the classification stage of AMC, including artificial neural networks (ANN), k-nearest neighbors (KNN), and support vector machines (SVM) \cite{Abdel13}. 

Recently, numerous machine learning and deep learning-based AMC methods \cite{Kim14, Karra15, Xie16, Bahloul17, Shi18, Hu19, Wang20, Huynh21, Zhao22, Li23, Nie24, Huang25, Lee26, Wang27, Wang28, Shah29, Salama30} have been developed for the classification of SISO and MIMO signals. These methods, as demonstrated in recent surveys \cite{Jdid31, Huynh32, Peng33, Zhang34, Zhou35, Zhu36}, show promise for achieving near-optimal performance with acceptable computational complexity. Some of these methods use various technologies to eliminate interference in complex MIMO signals and enhance classification performance. For instance, in \cite{Wang27, Wang28} authors propose a convolutional neural network-based zero-forcing equalization AMC method for MIMO systems. However, it is worth noting that the perfect elimination of MIMO signal interference cannot be achieved in real communication systems, and therefore the classification performance is decreased. 

In addition, authors in \cite{Zhang37, Zhang38, Wang39, Wei40} have proposed AMC algorithms for the classification of communication signals based on information entropy. Entropy, which measures the uncertainty of signal distribution and represents the complexity degree of the signal, plays a key role in these methods. Although these methods are competitive among other approaches, feature extraction using mutual information (MI) can be preferable over Shannon entropy, since MI considers the dependencies between two or more variables. Unlike entropy, which primarily focuses on the uncertainty of individual variables, MI measures how much information about one variable reduces uncertainty about another \cite{Cover41, Archer42, Press43}. This nuanced perspective is particularly valuable for extracting higher-level and interesting features from the data.

In this work, we propose a new AMC algorithm designed for MIMO system based on the mutual information feature extraction. Our algorithm stands out by specifically extracting features using IQ constellation diagrams of MIMO signals, a unique approach that has not been explored before. Furthermore, this method does not require the elimination of MIMO signal interference and a large amount of training data.  

The rest of the paper is organized as follows. System model, data set generation and experimental setup are introduced in section II. In Section III, MI background, the proposed AMC method, and SVM classifier are proposed. Section IV contains simulation results that demonstrate the correctness and validity of our method. Finally, conclusions are given in Section V.

\section{MIMO SYSTEM MODEL, DATA SET GENERATION AND EXPERIMENTAL SETUP}

\subsection{MIMO SYSTEM MODEL}

In this paper, we consider a MIMO system that consists of $N_s$ transmitting antennas and $N_r$ receiving antennas ($N_r$$\geq$$N_s$). Assuming that the MIMO channel is an uncorrelated Rayleigh flat-fading and time-invariant channel \cite{Wang28}, the received signal at the $n$-th sampling moment can be expressed as:
\begin{equation}
r(n)=Hs(n)+g(n),\label{eq1}
\end{equation}
where $H$ denotes $N_r$×$N_s$ dimensional MIMO channel matrix, and it follows circular symmetric complex normal distribution; $r(n)$ = $[r_1(n), r_{N_r} (n)]^T$ is the received signal vector with a size $N_r$×1; $s(n)$ = $[s_1(n), s_{N_t} (n)]^T$ is the transmitted signal vector with a size $N_t$×1; and $g(n)$ = $[g_1(n), g_{N_r} (n)]^T$  is  the additive white Gaussian noise vector with a size $N_s$ × 1 \cite{Wang2,Zhou3, Nie24}.

\subsection{Data set generation}

\begin{figure}[htbp!]
    \includegraphics[width=1\linewidth]{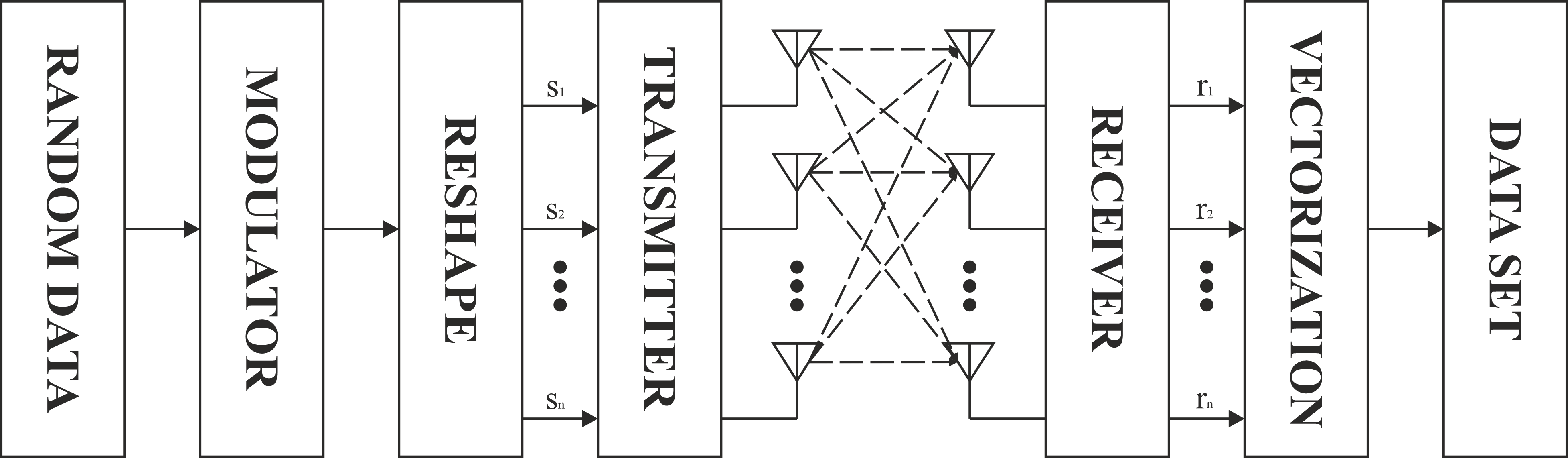}
    \caption{The process of dataset generation}
    \label{fig:1}
\end{figure}

Here we consider the system model of the complex multi- antenna, and the data set generating process is presented in Fig. \ref{fig:1}. Particularly, random data are modulated using various modulation types, such as binary phase shift keying (BPSK), quadrature phase shift keying (QPSK), 8-phase shift keying (8PSK), 16-quadrature amplitude modulation (16QAM), 64-quadrature amplitude modulation (64QAM). The modulation signal vector, represented as $S$, with size 1 × $N$ (where $N$ is the number of symbols, and $N$ = 256, 512, 1024). To ensure a fair comparison, $S$ is normalized with unit power, i. e., $\parallel S\parallel _2^2$ = 1. After normalization, $S$ is reshaped into a $N_s$ × $N/N_s$  matrix, represented as $[s_1, s_2, \ldots , s_{N_s}]$, here $s_i$ = $[s_i (1), s_i (2), \ldots ,s_i (N/N_s)$], $i\in [1, N_s];$ - is the transmitted signal vector at the $i$-th antenna.

The received signal vector at the $j$-th receive antenna passing through the MIMO channel, is denoted as $[r_1, r_2,\ldots, r_{N_s}]$, with size $N_r$ × $N/N_s$, $r_j$ = $[r_j(1), r_j(2),\ldots, r_j(N/N_s)], j\in [1, N_r]$.  Next, the received signal matrix can be vectorized into a 1 × $N$ vector $r$. Test samples and training data are retrieved from $r$. Specifically, the real and imaginary parts of $r:\mathcal{R}(r)$ and $\mathcal{I} (r)$ are separated and combined into a 2 × $N$ matrix $[\mathcal{R}(r)$ and $\mathcal{I} (r)]$, which is a sample for training and testing. It should be noted that 6,000 samples are prepared for training and 4,000 samples for testing for each SNR value \cite{Huang25, Wang27}.

\subsection{Experimental setup}

To evaluate the effectiveness and reliability of the proposed AMC method in real communication scenarios, we conducted a series of experiments using an integrated software and hardware complex for data collection of IQ constellation diagrams. The software component of this complex is implemented in the Matlab+Simulink program, while the hardware component comprises the Zedboard+AD9361 (Fig. 2). The Zedboard, connected to the PC via an Ethernet cable, provides communication between the software component and the RF module AD9361. The AD9361 RF module is compulsory as it provides wireless communication between receiver and transmitter. 

\begin{figure}[htbp!]
    \includegraphics[width=1\linewidth]{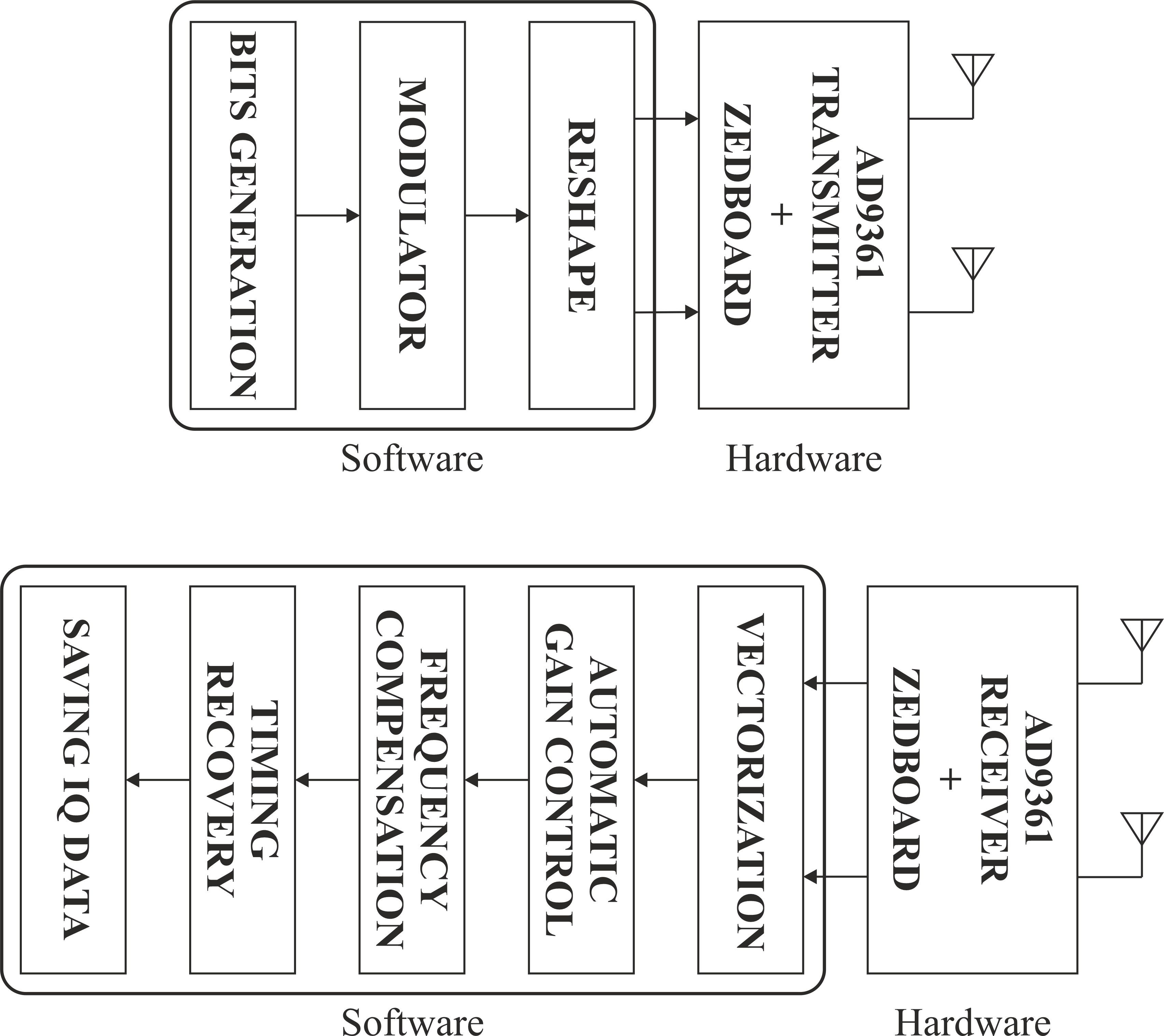}
    \caption{Transmitter and receiver block diagram }
    \label{fig:2}
\end{figure}

The transmitting unit of software part consists of a random bit generator, a modulator, a reshape demultiplexer device, while the hardware part comprises AD9361 with 2 transmitting antennas.

The receiving unit of the software part accomplishes multiplexing for vectorization and signal processing for IQ constellation diagrams. Signal processing, in turn, includes tasks such as automatic gain control (AGC), frequency compensation, and timing recovery. The hardware part of this receiving unit consists of AD9361 with 2 antennas (Fig 2). IQ constellation diagram values were saved and used for testing the proposed AMC algorithm. Fig. 3 below illustrates the hardware part of the experimental setup designed for receiving and transmitting signals in MIMO channels.

\begin{figure}[htbp!]
    \includegraphics[width=1\linewidth]{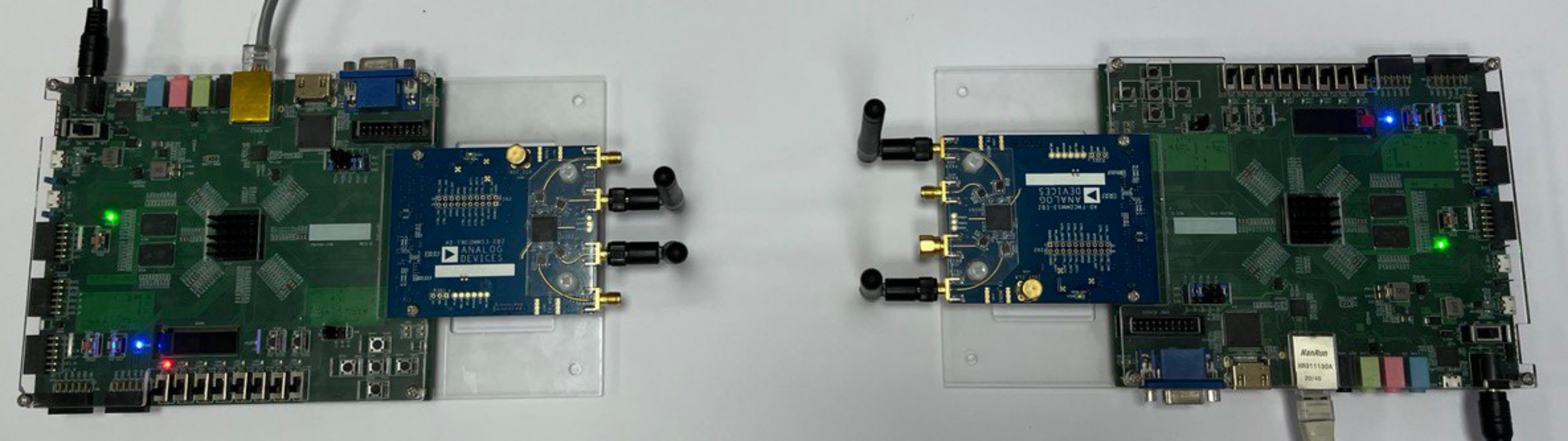}
    \caption{The hardware part of the experimental setup}
    \label{fig:3}
\end{figure}

The configuration of the AD9361 RF module was performed using Simulink, and the following parameters were selected: a frequency of 2.4 GHz, a baseband sample rate of 520.841 kHz. The transmitter and receiver gain parameters were adjusted to control the Signal-to-Noise Ratio (SNR) value. 
The experimental results confirm our theoretical data and serve as a qualitative addition to this research.

\section{The proposed AMC method based on mutual information feature extraction}

In this section, we describe the structure of the proposed AMC method, which is shown in Fig 4. The method consists of two main parts: feature extraction based on mutual information $I(X;Y)$ and SVM classifier. To make it easier to understand, we introduce the section from three aspects: MI background, detailed explanation of the proposed method and SVM classifier.  

\begin{figure}[htbp!]
    \includegraphics[width=1\linewidth]{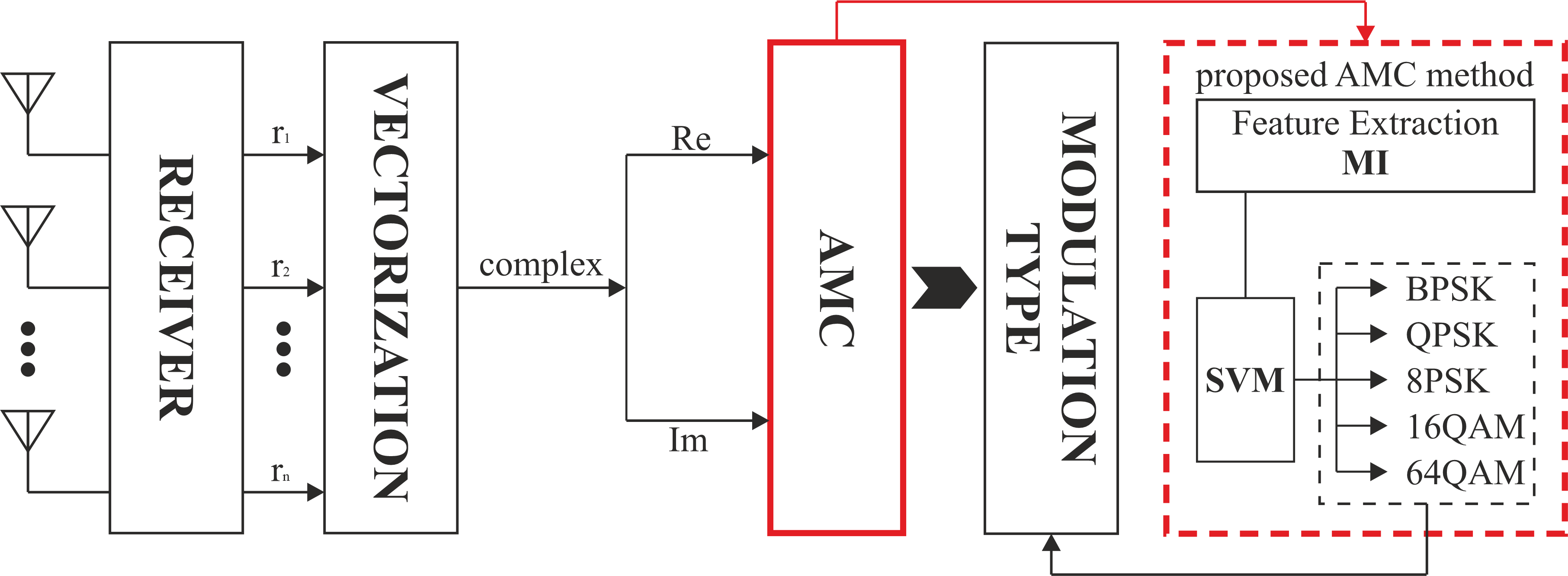}
    \caption{The structure of the proposed AMC method}
    \label{fig:4}
\end{figure}

\subsection{MI background}

In probability theory and information theory, mutual information is a basic concept that measures the statistical dependence between two random variables. Specifically, it represents the amount of information learned about one random variable by observing another. Mutual information is closely related to the entropy because knowing information about one random variable can decrease the level of uncertainty of another variable. Therefore, a high value of $I(X;Y)$ shows a substantial reduction of uncertainty, while a low value signifies a minor reduction. If the mutual information between two random variables is zero, it indicates that they are independent of each other \cite{Cover41, Archer42}. 

The mutual information between two discrete random variables $X$ and $Y$ is defined as a double sum:
\begin{equation}
    I(X;Y)= \sum_{j\in Y}^{N}\sum_{i\in X}^{M}P_{(i,j)}(x,y)log_2 \left(\frac{P_{(i,j)}(x,y)}{P_i(x)P_j(y)}\right),
 \label{eq2}
\end{equation}
where $N$ is the number of possible values for $X$ and $M$ is the number of possible values for $Y,P_{(i,j)}(x,y)$ is the joint probability density function, $P_i(x)$ and $P_j(y)$  denote the marginal probability density function of $X$ and $Y$ \cite{Cover41,Archer42,Press43, Batina44, Veyrat45}.

As explained before, there is a strong relationship between mutual information and entropy. The following formula illustrates this relation:
\begin{equation} 
I(X;Y ) = H(X)-H(X/Y),\label{eq3}
\end{equation}
where $H(X)$, $H(X/Y)$ are the marginal and conditional entropies. To comprehend the meaning of $I(X;Y)$, it is necessary to explain both entropy and conditional entropy. Entropy ($H$) quantifies the degree of expected uncertainty of a random variable, and it can be expressed as follows \cite{Batina44}:  

\begin{equation}
    H(X) = -\sum_{i} P_i (x) log_2 P_i (x),\label{eq4}
 \end{equation}
where $P_i$ is the probability of event  $x_i$. The conditional entropy quantifies the amount of uncertainty that a random variable $X$ has when the value of $Y$ is given, and it can be defined using the following formula \cite{Veyrat45}:

\begin{equation}
H(X/Y)=-\sum_{i=1}^{N}\sum_{j=1}^{M}P_{(i,j)}(x,y)log_2(P_{(i,j)}(x/y)), \label{eq5}     
\end{equation}
where $N$ is the number of possible values for $X$ and $M$ is the number of possible values for $Y, P_{(i,j)}(x,y)$  denotes the joint probability,and $P_{(i,j)}(x/y)$  denotes the conditional probability. 

\subsection{The proposed method}
The proposed method involves the extraction of features based on mutual information (MI) using a histogram of IQ diagram. The process begins with constructing an IQ diagram that depicts the complex values of modulated signals. The in-phase component of the received signal is chosen as variable $X$, while the quadrature component is denoted as variable $Y$. Then, to calculate MI we built histogram (Fig. 5) of $XY$ plane divided into bins of equal size (\(\Delta X \times \Delta Y\)) with coordinates $i,j$.  By counting the number $k_{(i,j)}$ of samples in bin $(i,j)$, MI can be calculated using the following formula: 

\begin{equation}
    I(X;Y) = \sum_{j=1}^{k_Y} \sum_{i=1}^{k_X} \left( \frac{k_{ij}}{N} \right) \log \left( \frac{k_{ij} N}{k_i k_j} \right)
 \label{eq6}
\end{equation}

To implement the proposed algorithm, the following steps must be performed:

1. Construct IQ diagram.

2. Build histogram of two-dimensional IQ diagram.

3. MI calculation according to equation (6).

4. In the final step, apply an SVM classifier to identify the type of modulation.

In Fig. 5, the histogram of the IQ diagrams shows significant differences corresponding to the type of modulation. The point density decreases as the number of symbols increases.

\begin{figure}[htbp!]
    \includegraphics[width=1\linewidth]{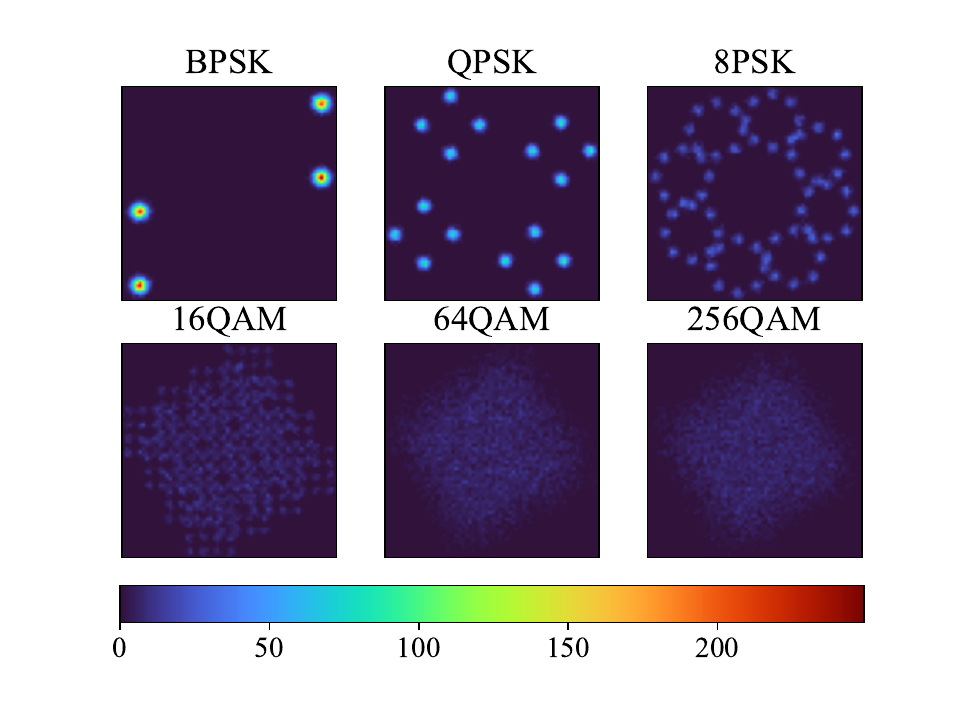}
    \caption{Histogram of IQ diagrams of four modulation schemes at SNR=25}
    \label{fig:5}
\end{figure}

In the literature, several rules have been proposed to select the number of bins ($b$), such as the choice of the square root (where \( b = \sqrt{N} \)), the Sturges formula (where \( b = 1 + \log_2{N} \)), the Rice rule (where \( b = 2N^{1/3} \)) \cite{Balzanella46}. The comparison of these rules is given in the results.

\subsection{SVM classifier}

Various types of classifiers have been used in the classification stage of AMC until those days. In this work, we evaluated the performance of selected classifiers based on their Probability Correct Classification (Table 1) and chose the SVM classifier.

Support Vector Machine (SVM) is a supervised machine learning algorithm developed by Vapnik and Chervonenkis based on statistical learning theory \cite{Awad47}. It is useful for solving small-sample learning problems and has been successfully applied in various fields, including pattern recognition, regression analysis, and density estimation. SVM using kernel functions maps the input data into a high-dimensional feature space, where a nonlinear classification problem can be treated as a linear one.

 By applying the rule of structural risk minimization, SVM aims to not only reduce classification errors but also enhance the generalization ability of classifiers. Compared to traditional Neural Network algorithms, SVM is advantageous due to its simple structure, high success rates, better generalization, and small-sample-problem-solving ability. Detailed mathematical theories of SVM are available in references \cite{Veisi48} and \cite{Ding49}. 

\section{Results and discussion}
This section presents simulation results to evaluate the performance of the suggested modulation classification scheme. We analyze a MIMO communication system with $N_s = 2$ transmitting antennas and $N_r = 2$ receiving antennas. A channel with relay fading and AWGN noise is assumed. Six modulation schemes are considered: BPSK, QPSK, 8PSK, 16QAM, 64QAM. The extraction of signal features using mutual information is calculated by averaging 100 simulations using the Monte Carlo method (Fig. 6).

\begin{figure}[htbp!]
    \includegraphics[width=1\linewidth]{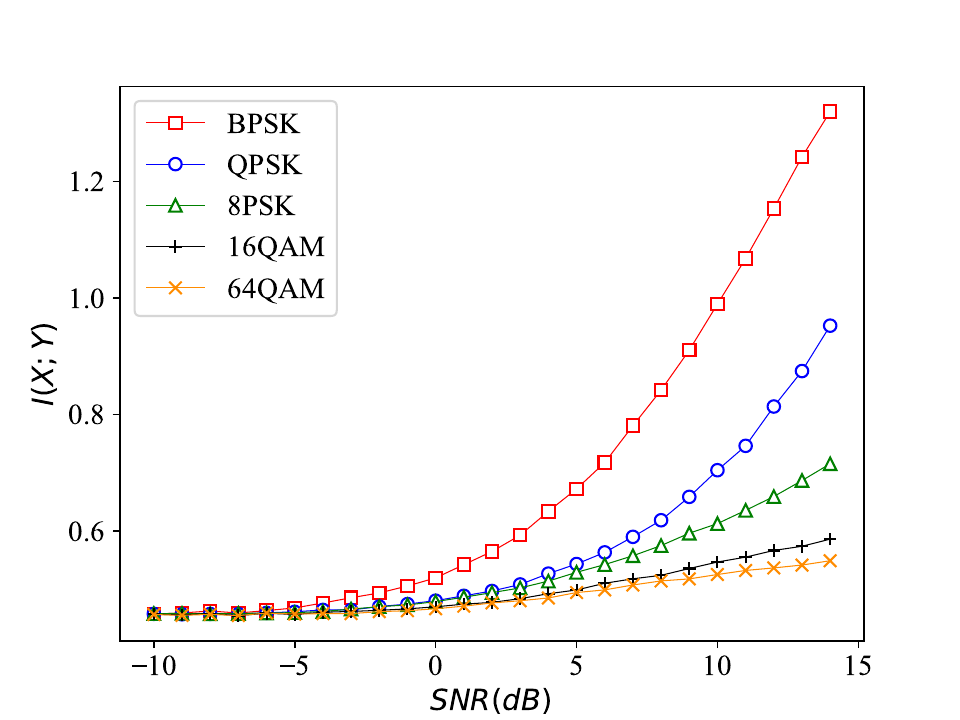}
    \caption{Feature extraction based on mutual information under different SNR}
    \label{fig:6}
\end{figure}

According to Fig. 6, as the number of symbols on the constellation increases, the distance between the points increases, which leads to a decrease in the density of points on the constellation and, consequently, to a decrease in mutual information. To calculate the probability of correct classification, the following formula is used:
\begin{equation}
    P_{cc} = N_{cc}/N_{rs}, 
\label{eq7}  
\end{equation}
where $N_{cc}$ is the number of correctly classified signals, $N_{rs}$ is the total number of received signals. Fig. 7 below shows a comparison of the existing rules for selecting bins.

\begin{figure}[htbp!]
    \includegraphics[width=1\linewidth]{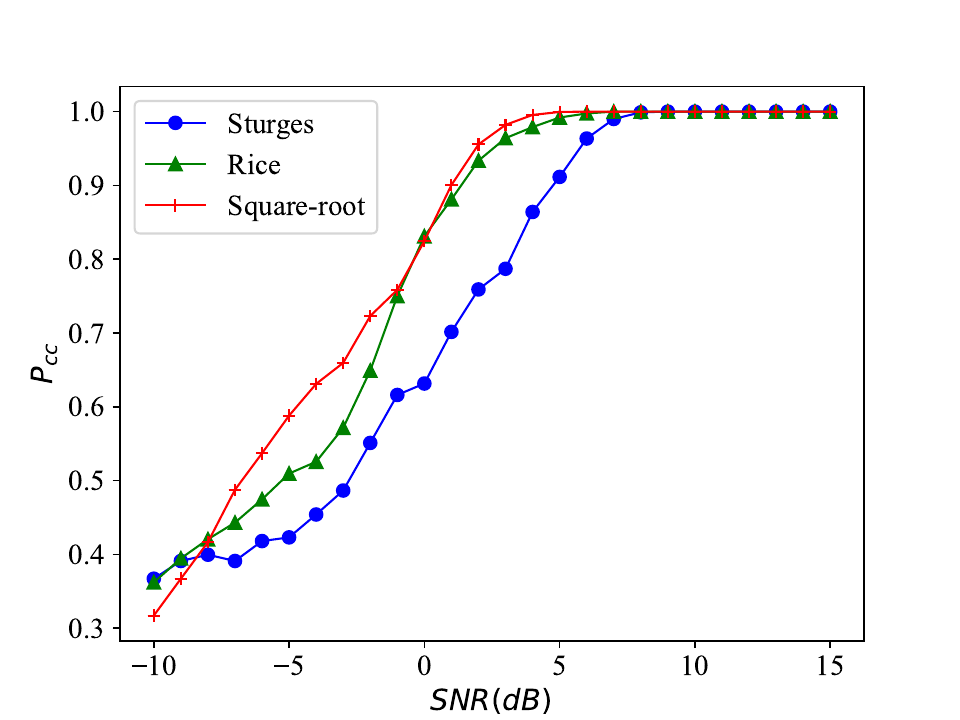}
    \caption{Comparison of empirical rules for selecting bins (N=1024).}
    \label{fig:7}
\end{figure}

According to Fig. 7, the results of the square root and Rice rule demonstrate better performance compared to the Sturges rule. This is because, from the perspective of the integral root-mean-square error, a larger number of bins enhances the accuracy of the density estimate. 

Also, for the proposed AMC algorithm, we assessed the average probability of correct classification across different classifiers at SNR from -10 to 15, and we also examined the simulation time of these classifiers (Table 1). Simulations were conducted using a computer equipped with an 11th Gen Intel(R) Core (TM) i5-11400F processor operating at 2.60 GHz and 16.0 GB of RAM. The probability of correct classification of the classifier is generally considered the most crucial indicator for evaluating a feature extraction algorithm. We use five classifiers, including KNN, SVM, Adaboost, Random Forest (RF) and MLP to classify the original feature set extracted by our proposed AMC method. These classifiers are evaluated using different methods for determining the number of bins in a histogram, including Sturges, Rice, and Square Root. The specific parameter for the KNN classifier was set as follows: the nearest neighbor number ($k$) was set to 5. For the SVM classifier, the kernel function was set as RBF kernel function. The Adaboost classifier's parameters were set with a depth of 10, a learning rate of 0.1, and 10 iterations. The Random Forest classifier had a depth of 9, a learning rate of 0.1, and 10 iterations. The MLP classifier was configured with a learning rate of 0.1, and 10 iterations. According to the data set, the simulation time and average probability of correct classification of different classifiers are calculated. 

\begin{table}[!ht]
\centering
\caption{Performance Comparison of Classifiers for the Proposed AMC Algorithm}
\setlength\tabcolsep{4pt}
\renewcommand{\arraystretch}{1.2} 
\small
\begin{tabular}{|l|cc|cc|cc|}
\hline
& \multicolumn{2}{c|}{Sturges} & \multicolumn{2}{c|}{Rice} & \multicolumn{2}{c|}{Square-root} \\
\cline{2-7}
& \( P_{cc} \) & t (s)& \( P_{cc} \) & t (s) & \( P_{cc} \) & t (s)\\
\hline
SVM & 79.21\% & 0.56 & 84.79\% & 0.48 & 85.24\% & 0.41 \\
Adaboost & 62.60\% & 0.25 & 66.29\% & 0.11 & 68.72\% & 0.25 \\
MLP & 78.92\% & 0.37 & 83.89\% & 0.34 & 84.37\% & 0.62 \\
RF & 78.84\% & 0.09 & 81.41\% & 0.19 & 82.31\% & 0.19 \\
KNN & 80.25\% & 0.10 & 83.85\% & 0.09 & 84.06\% & 0.09 \\
\hline
\end{tabular}
\end{table}

Table 1 reveals that when applying the square root rule,
SVM, MLP, and KNN classifiers achieve the highest average probability of correct classification compared to the Sturges and Rice methods. Specifically, SVM classifier leads with the highest average probability of correct classification, closely followed by MLP and KNN. On the other hand, Adaboost and RF classifiers exhibit lower average probabilities of correct classification across all methods. In terms of simulation times, KNN stands out as the most efficient among the three top-performing classifiers, demonstrating shorter times compared to SVM and MLP. Based on these results, in this paper, we favor the use of the SVM classifier because of its highest average probability of correct classification, prioritizing the sensitivity of feature extraction using MI over the simulation time of the classifiers.

Fig. 8 presents the simulation results of our proposed AMC method with a MIMO configuration of two transmit and two receive antennas (2x2). Fig. 8 shows the $P_{cc}$ curves against different SNR values for various classifiers, including SVM and KNN classifiers. The proposed KNN and SVM classifiers based on mutual information is superior to the KNN and SVM classifier based on HOC, given in \cite{Tayakout50}.  

\begin{figure}[htbp!]
    \includegraphics[width=1\linewidth]{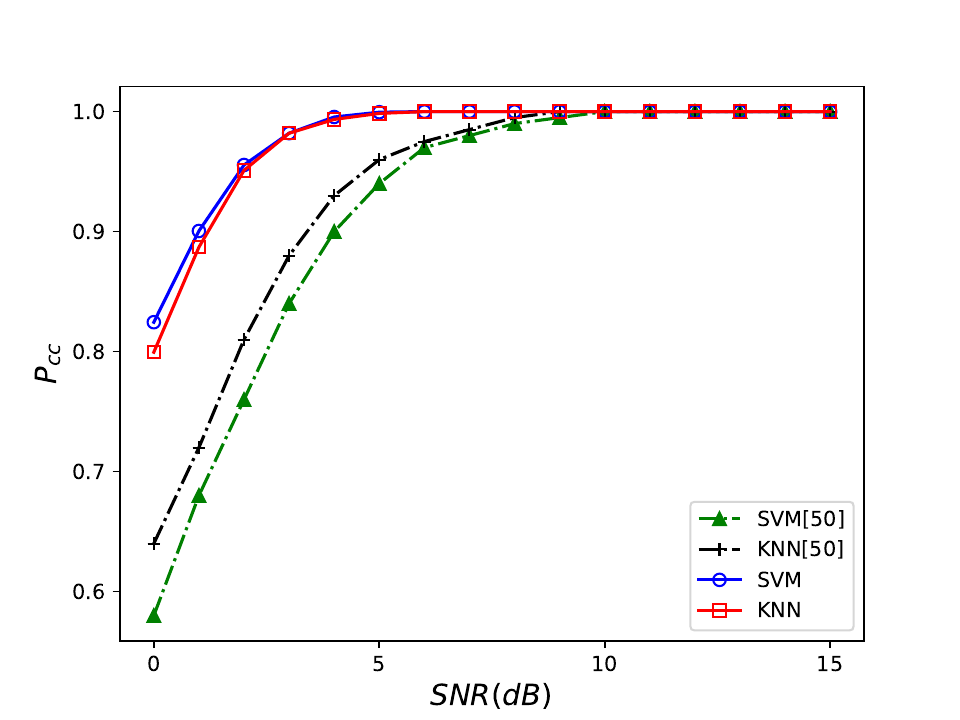}
    \caption{Performance Comparison between MI-based and HOC-based Classification Methods}
    \label{fig:8}
\end{figure}

Our proposed future based classifier outperforms classifier based on HOC and achieves perfect classification at 5 dB SNR, whereas the compared classifier attains 100\% classification at SNR of 13 dB.  We conducted a comparison of the simulation times between our proposed method and the HOC-based approach. The algorithm execution time for our method was 0.1 seconds for KNN and 0.56 seconds for SVM, slightly longer than the HOC-based method, which recorded times of 0.53 seconds for SVM and 0.07 seconds for KNN.

Figure 9 shows the effect of N on the performance of our suggested method under different SNR values, where N = 256, 512, 1024. Figure 9 shows that the probability of correct classification increases with the growth of SNR, and then stabilizes. The five modulation types BPSK, QPSK, 8PSK, 16QAM, 64QAM are classified correctly with 100\% probability when the SNR is about 5 dB. According to Figure 9, the probability of correct $P_{cc}$ classification from the number of points varies with a small error, since mutual information is insensitive to the size of the data. Figure 10 below illustrates a comparison of the probability of correct classification for each modulation scheme relative to the other considered modulations.

\begin{figure}[b]
    \includegraphics[width=1\linewidth]{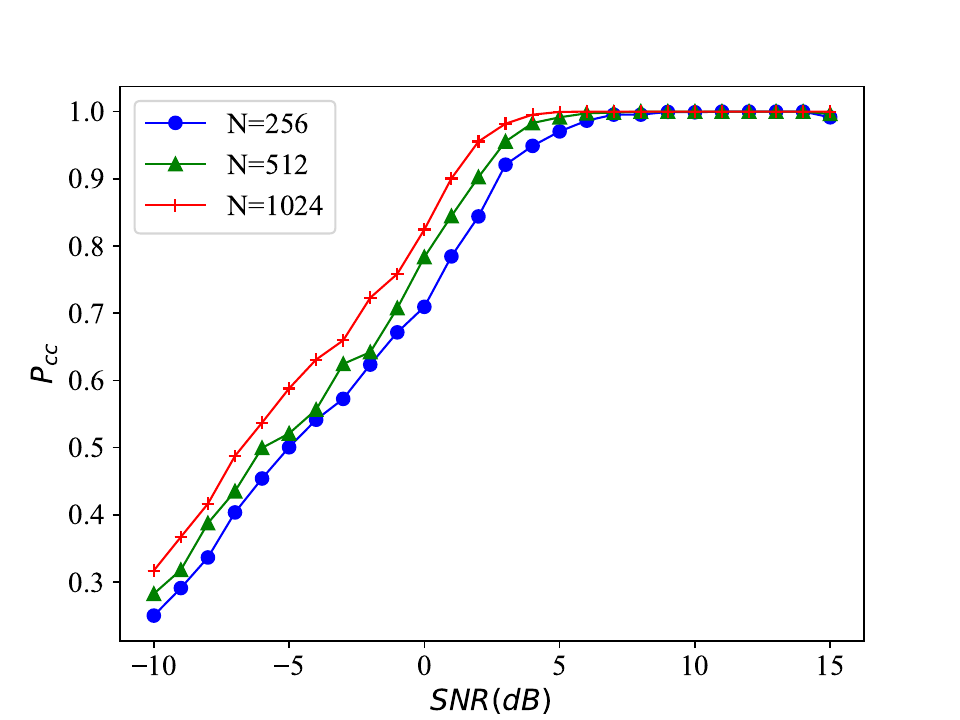}
    \caption{Average probability of correct classification for different N}
    \label{fig:9}
\end{figure}

\begin{figure}[htbp!]
    \includegraphics[width=1\linewidth]{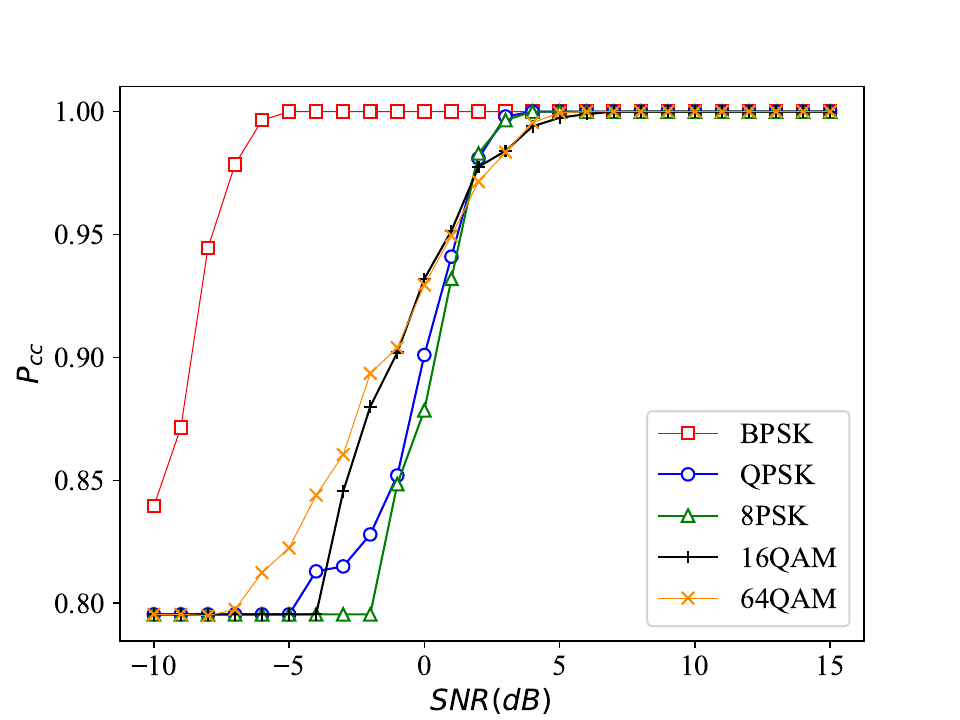}
    \caption{Probability of correct classification ($P_{cc}$) among various modulation schemes (N=1024)}
    \label{fig:10}
\end{figure}

In general, the classification accuracy of QAM signals is determined better in comparison with PSK signals. Since with an increase in the modulation order of QAM signals, the points on the constellation are distributed more evenly. It is revealed that BPSK signals can be well identified using the classifier proposed by us even at low SNR = -5 dB. The QPSK and 8PSK signals can be correctly recognized at SNR above 3 dB. Perfect 16QAM and 64 QAM detection is achieved at 5 dB SNR. Figure 11 shows the probabilities of correct classification of BPSK and QPSK signals obtained using the experimental setup.

\begin{figure}[htbp!]
    \includegraphics[width=1\linewidth]{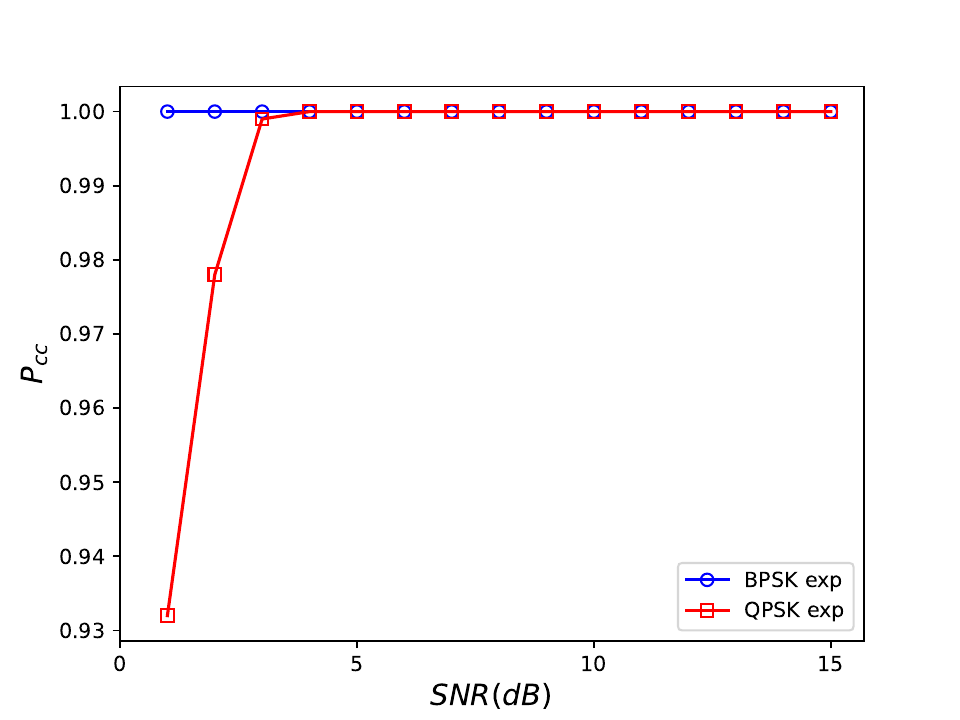}
    \caption{Average probability of correct classification of experimental data}
    \label{fig:11}
\end{figure}

As shown in Figure 11, the QPSK signal can be correctly recognized at SNR above 3 dB, as in the theoretical results.

\section{Conclusion}
\label{sec:conclusion}

In this article, we proposed a new classifier based on the extraction of mutual information features for recognizing the type of modulation schemes with low SNR for MIMO systems. This method does not require the elimination of MIMO signal interference and a large amount of training data, since our method is more resistant to noise uncertainty and spatial correlation in MIMO systems.

In addition, it is worth noting that feature extraction methods based on mutual information were not applied in AMC algorithms for MIMO systems. Our approach proves its effectiveness by using mutual information that considers the interdependencies between variables. The modeling and experimental results showed that our algorithm can work properly even at low SNR with a small number of observation samples, thereby representing a promising advancement in the recognition of modulation schemes within MIMO systems.

\section{Acknowledgement}
We thank the anonymous referee for constructive comments that helped to improve this article.
This research was funded by the Committee of Science of the Ministry of Science and Higher Education of the Republic of Kazakhstan (Grant No. AP14872061).

\bibliographystyle{unsrt}
\bibliography{refs.bib}

\begin{IEEEbiography}[{\includegraphics[width=1in,height=1.25 in,clip,keepaspectratio]{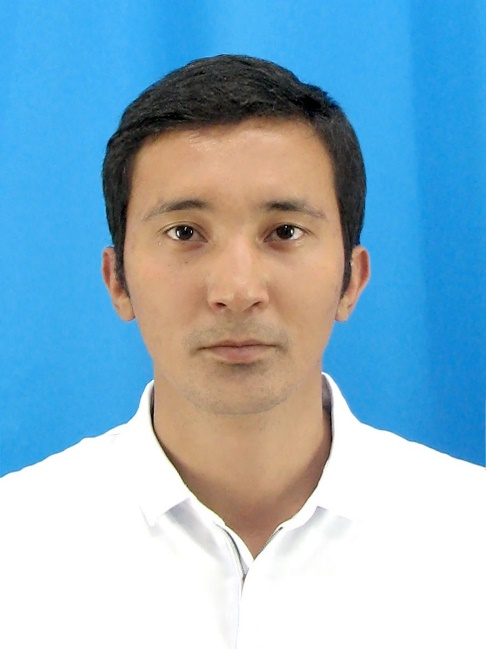}}]{N. Ussipov Author} received the master’s degree in engineering sciences from Taraz State University named after M. Kh. Dulaty, Taraz, Kazakhstan, in 2016.  He is currently a senior researcher and chief programmer at the Department of Solid State Physics and Nonlinear Physics  of the Al-Farabi Kazakh National University, where he is pursuing the Ph.D. degree. His current research interests include signal modulation classification, signal processing, network implementation, network information theory, routing algorithms. 
\end{IEEEbiography}

\begin{IEEEbiography}[{\includegraphics[width=1in,height=1.25in,clip,keepaspectratio]{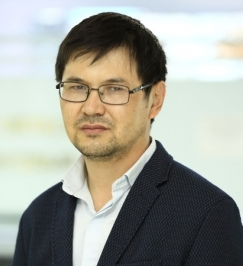}}]{S. Akhtanov   Author} received the Ph.D. degree in physics from Al-Farabi Kazakh National University, Almaty, Kazakhstan, in 2019. He is currently a senior researcher at the Department of Solid State Physics and Nonlinear Physics of the Al-Farabi Kazakh National University. His current research interests include signal modulation classification, signal processing, network implementation, and optimization algorithms in wireless networks. 
\end{IEEEbiography}

\begin{IEEEbiography}[{\includegraphics[width=1in,height=1.25in,clip,keepaspectratio]{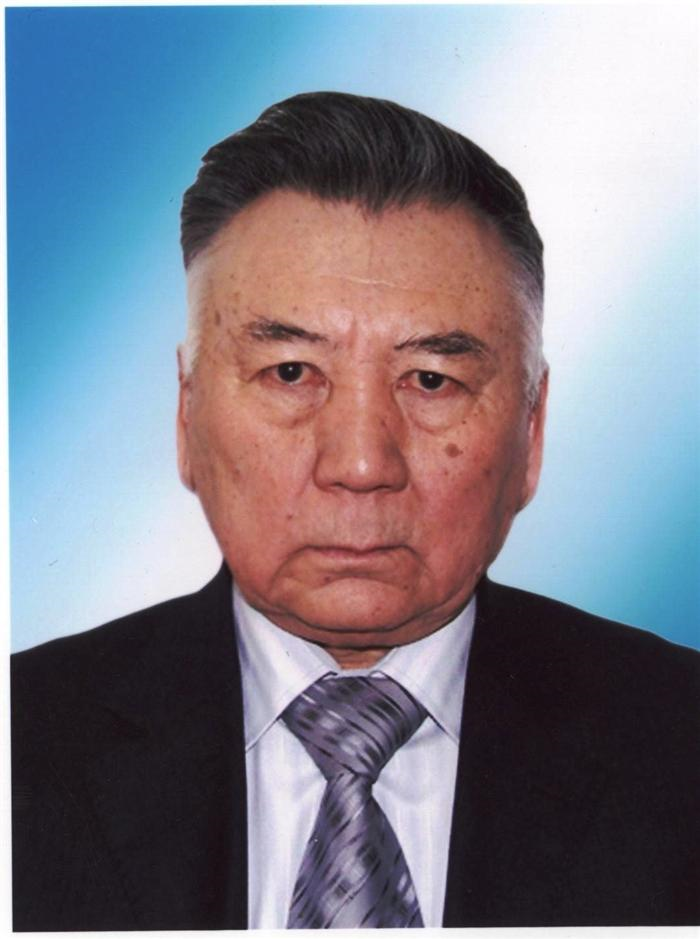}}]{Z. Zhanabayev  Author} received the degree of Doctor of Physical and Mathematical Sciences from Al-Farabi Kazakh National University, Almaty, Kazakhstan, in 1996. In 1997, he received the title of professor from Al-Farabi Kazakh National University. Under his leadership, 9 candidate dissertations, 3 doctoral dissertations, 5 doctoral dissertations of international level (PhD) were defended. Professor Zhanabaev Z.Zh. is the initiator of the development of a new scientific direction in Kazakhstan ‘nonlinear physics and synergetics’. Professor Z.Zh. Zhanabayev has written 2 monographs, 12 textbooks, more than 300 scientific articles on modern areas of science.
\end{IEEEbiography}

\begin{IEEEbiography}[{\includegraphics[width=1in,height=1.25in,clip,keepaspectratio]{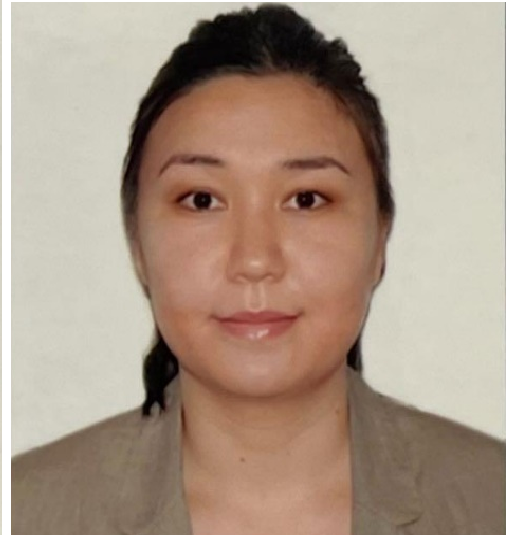}}]{D. Turlykozhayeva  Author} received the master’s degree in physics from Tomsk Polytechnical University, Tomsk, Russia, in 2017.  She is currently a senior researcher and chief editor at the Department of Solid State Physics and Nonlinear Physics of the Al-Farabi Kazakh National University, where she is pursuing the Ph.D. degree. Her current research interests include signal modulation classification, signal processing, network implementation, network theory, routing of wireless networks.
\end{IEEEbiography}

\begin{IEEEbiography}[{\includegraphics[width=1in,height=1.25in,clip,keepaspectratio]{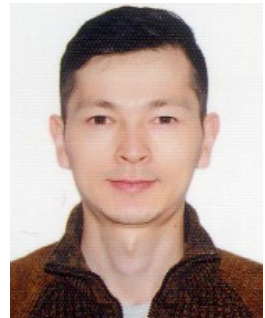}}]{B. Karibayev  Author} received the Ph.D. degree in radio engineering, electronics and telecommunication from Al-Farabi Kazakh National University, Almaty, Kazakhstan, in 2019. He is currently a senior researcher at the Department of Telecommunication Engineering of the Almaty University of Power Engineering and Telecommunications named after Gumarbek Daukeyev. His current research interests include signal modulation classification, signal processing, computer modeling, MIMO antennas.
\end{IEEEbiography}

\begin{IEEEbiography}[{\includegraphics[width=1in,height=1.25in,clip,keepaspectratio]{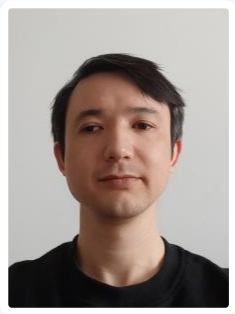}}]{T. Namazbayev Author} received the master’s degree in radio engineering, electronics, and telecommunication from Al-Farabi Kazakh National University, Almaty, Kazakhstan, in 2019.  He is currently a senior researcher at the Department of Solid State Physics and Nonlinear Physics of the Al-Farabi Kazakh National University. His current research interests include signal modulation classification, signal processing, computer modeling, Mimo antennas.
\end{IEEEbiography}

\begin{IEEEbiography}[{\includegraphics[width=1in,height=1.25in,clip,keepaspectratio]{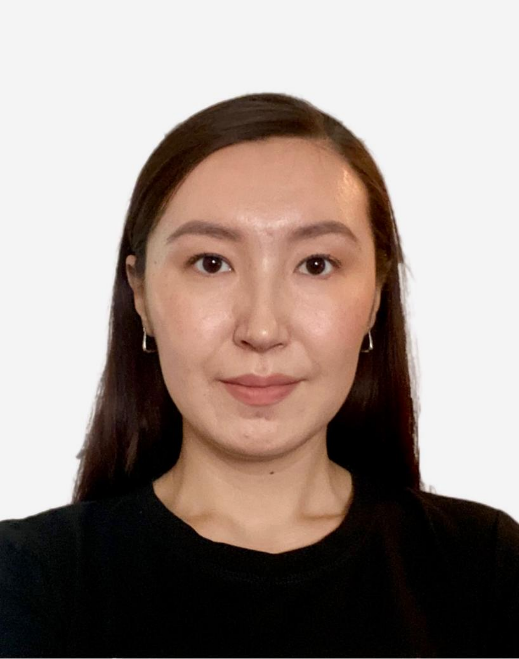}}]{D. Almen  Author} received the master’s degree in radio engineering, electronics and telecommunication from Al-Farabi Kazakh National University, Almaty, Kazakhstan, in 2021.  She is currently a researcher at the Department of Solid State Physics and Nonlinear Physics of the Al-Farabi Kazakh National University. Her current research interests include signal modulation classification, signal processing, wireless networks theory, wireless communication. 
\end{IEEEbiography}

\begin{IEEEbiography}[{\includegraphics[width=1in,height=1.25in,clip,keepaspectratio]{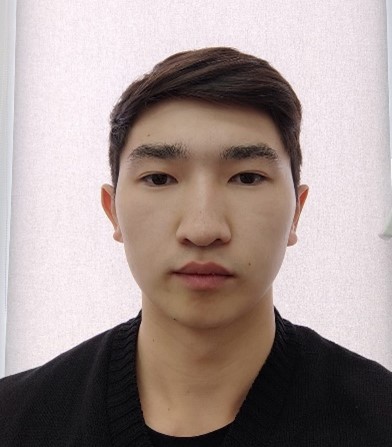}}]{A. Akhmetali  Author} student of Al-Farabi Kazakh National University, Almaty, Kazakhstan. He is currently a junior researcher at the Department of Solid State Physics and Nonlinear Physics of the Al-Farabi Kazakh National University. His current research interests include signal processing, network information theory, network implementation, routing algorithms.
\end{IEEEbiography}

\begin{IEEEbiography}[{\includegraphics[width=1in,height=1.25in,clip,keepaspectratio]{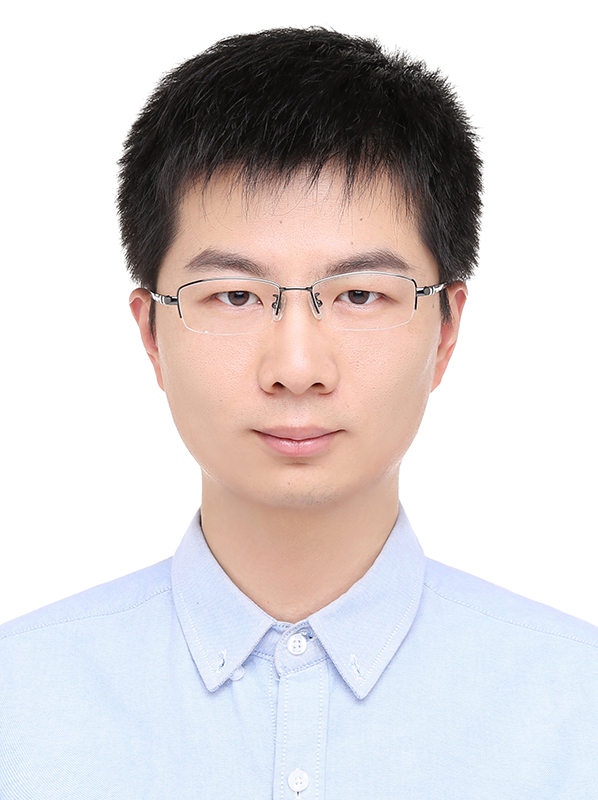}}]{Xiao Tang Author (Member, IEEE)} received the B.S. degree in information engineering (Elite Class Named After Tsien Hsue-shen) and Ph.D. degree in information and communication engineering from Xi’an Jiaotong University, Xi’an, China, in 2011 and 2018, respectively. He is currently with the Department of Communication Engineering, Northwestern Polytechnical University, Xi’an, China. His research interests include wireless communications and networking, game theory, and physical layer security.
\end{IEEEbiography}

\EOD

\end{document}